\documentclass[letterpaper,10pt,conference]{ieeeconf}
\usepackage{cite}
\usepackage{amsmath,amssymb}
\usepackage{graphicx}
\usepackage{booktabs}
\usepackage{multirow}
\usepackage{hyperref}
\usepackage{color}
\usepackage{soul}
\usepackage{tikz}
\usepackage{xcolor}
\makeatletter
\let\NAT@parse\undefined
\makeatother
\usepackage{hyperref}
\usetikzlibrary{arrows.meta}

\makeatletter
\patchcmd{\@makecaption}{{\footnotesize #1}\\{\footnotesize\scshape #2}}{{\footnotesize #1.~~#2\par}}{}{}
\makeatother

\definecolor{diagramblue}{RGB}{0,100,166}
\definecolor{testperiod}{RGB}{205,55,0}
\definecolor{evaluationperiod}{RGB}{255,185,15}

\IEEEoverridecommandlockouts
\newcommand{\casename}{\texttt{bestest\_\allowbreak hydronic\_\allowbreak heat\_\allowbreak pump}}

\title{\LARGE\bfseries Comfort as a Constraint or as a Cost in Reinforcement Learning and Model Predictive Building Control}

\author{%
    Joram Liebeskind$^{1}$, %
    Valentin Bhend$^{1}$, %
    Varsha N. Behrunani$^{2}$, %
    Alisa Rupenyan$^{1}$ %
    \thanks{$^1$ Centre for Artificial Intelligence, Zurich University of Applied Sciences, Switzerland. liej@zhaw.ch, bhed@zhaw.ch, rupn@zhaw.ch}
    \thanks{$^2$ Belimo Automation AG, varsha.behrunani@belimo.ch}
}

\begin{document}
\maketitle
\thispagestyle{empty}
\pagestyle{empty}

\begin{abstract}
A building controller that keeps a zone within comfort bounds at minimum energy can treat the bounds as a soft constraint, by adding the weighted violation to its cost, or as a hard constraint.
We analyze the effect of the hard and the soft formulation of the same problem and isolate the effect of the formulation from the effect of the algorithm.

To this end, we measure the difference in energy use between the two formulations on the BOPTEST \casename{} case, in model predictive control (MPC) and in reinforcement learning (RL).
The soft formulations are a weighted MPC and weighted SAC and DDPG learners, each at twelve weights.
The hard formulations are an MPC with an exact penalty and the Lagrangian learners SAC-Lagrangian and primal-dual DDPG. We find that the hard formulations use more energy than the soft formulations at the largest weight, but violate the comfort bounds less. For a large weight, the soft formulation approaches the behavior of the hard formulation.
The frontiers of the energy--comfort trade-off of the three weighted controllers lie close to each other, and the weight selects an operating point on this frontier.
Under the soft formulation, the three algorithms reach very similar operating points, although at somewhat different weights. For this HVAC problem, the formulation and the weight thus affect the energy use more than the choice of algorithm.

\end{abstract}

\section{INTRODUCTION}

Building climate control with an electric heat pump is defined by the minimization of the electrical energy of the heat pump, subject to comfort bounds on the zone temperature~\cite{oldewurtelUseModelPredictive2012,drgonaAllYouNeed2020}.
The bounds, however, admit two formulations that define different controllers.

In a \emph{hard-constrained} formulation, the controller minimizes the energy subject to no violation of the bounds.
This controller is determined by the bounds, but the problem can become infeasible, and its behavior in that case has to be defined separately~\cite{scokaertFeasibilityIssuesLinear1999}. In a \emph{soft-constrained} formulation, the controller minimizes the energy plus a weighted penalty on the violation~\cite{kerriganSoftConstraintsExact2000}.
This problem is always feasible, but each weight gives a different operating point on the energy--comfort trade-off \cite{zhangConstrainedReinforcementLearning2026,sanchezConstrainedReinforcementLearning2024}.

How difficult the bounds are to hold depends on disturbances outside the controller's influence, the weather and the internal gains, and on the occupancy, which sets when the tighter bounds apply.
MPC solves an optimization problem over a model of the building and treats the comfort bounds as a constraint with a slack variable, whose weight determines whether the formulation is soft or hard. Even under the hard formulation, the comfort bounds hold on the building only as far as the model and the forecasts of the disturbances are accurate. Stochastic MPC replaces the bounds with chance constraints, at the cost of a model of the disturbance distribution~\cite{oldewurtelUseModelPredictive2012}.
A Lagrangian reinforcement learner enforces the bounds on average over the disturbances~\cite{altmanConstrainedMarkovDecision1999,tesslerRewardConstrainedPolicy2018}, and its Lagrange multiplier acts as a weight that is adapted during training.
Comparisons of the two methods, including on the common benchmark BOPTEST~\cite{blumBuildingOptimizationTesting2021}, fix one formulation per method~\cite{arroyoComparisonOptimalControl2022,zhangConstrainedReinforcementLearning2026}, which leaves open how much of the difference between the methods is due to the formulation.

This paper examines how the choice between the soft and the hard formulation of the comfort bounds affects a controller that minimizes energy within a comfort band. We quantify the cost of each formulation in energy and in comfort, and we examine whether the trade-off between the two changes predictably with the weight of the soft formulation, so that the weight can be used to select an operating point on the frontier.

The paper's contributions are as follows.
First, we compare the soft and the hard formulation of the comfort bounds in MPC and in RL.
We find that the hard formulation behaves in both methods as a soft formulation with a large weight.
Second, we map the energy–comfort trade-off over twelve weights in the interval $[0.02,2]$.
 The weight determines the energy use mainly between $w \approx 0.05$ and $0.4$.
For larger weights, the soft formulation approaches the hard formulation, and the energy needed to avoid one Kelvin-hour of comfort violation grows by about two orders of magnitude.
Third, we compare MPC and RL systematically on the high-fidelity BOPTEST \casename{} simulation. At the largest weight of the grid, the three algorithms differ by at most 2\,\% in energy, while the formulation changes the energy use by 7 to 33\,\%.
The formulation and the weight therefore affect the energy use more than the choice of algorithm.

\section{RELATED WORK}

Both MPC and RL can formulate comfort as a soft or as a hard constraint.
MPC is the established optimization-based method for building climate control~\cite{drgonaAllYouNeed2020,seraleModelPredictiveControl2018} and usually represents comfort by temperature bounds with a weighted slack.
An exact penalty, a slack weight above a problem-dependent threshold, returns the hard-constrained solution where one exists and the solution with the least violation otherwise~\cite{kerriganSoftConstraintsExact2000}.
Deep RL has been applied to HVAC control since~\cite{weiDeepReinforcementLearning2017}, among others for multi-zone residential control~\cite{duIntelligentMultizoneResidential2021}, occupant-centric HVAC and window control~\cite{liuOccupantcentricHVACWindow2024}, demand response~\cite{sanchezConstrainedReinforcementLearning2024} and heat pumps~\cite{zhangConstrainedReinforcementLearning2026}, see the review~\cite{wangReinforcementLearningBuilding2020}.
The reward is usually a weighted sum of energy and discomfort~\cite{duIntelligentMultizoneResidential2021}.
Constrained RL instead formulates a requirement as a Constrained Markov Decision Process (CMDP)~\cite{altmanConstrainedMarkovDecision1999}.
Lagrangian methods alternate a policy update with dual ascent on a multiplier~\cite{tesslerRewardConstrainedPolicy2018,liangAcceleratedPrimalDualPolicy2018}.
PID-Lagrangian methods damp the multiplier~\cite{stookeResponsiveSafetyReinforcement2020}, and other methods bound a worst-case cost~\cite{yangWCSACWorstCaseSoft2021}, solve the problem variationally~\cite{liuConstrainedVariationalPolicy2022} or project actions through a safety layer~\cite{ceustersAdaptiveSafetyLayer2023}, see~\cite{guReviewSafeReinforcement2024}.
For heat pumps, a constrained SAC with a log-barrier function has been applied~\cite{zhangConstrainedReinforcementLearning2026}.

The formulations, and the weight of the soft formulation, have been studied within one method at a time.
Within MPC, the formulation of the optimization problem and the relaxation of its constraints change energy use and comfort~\cite{huangUnveilingOverlookedAspects2024}, and formulations have been compared for an air-source heat pump with floor heating~\cite{zanettiPerformanceComparisonQuadratic2023}.
The allowed violation can also be specified as a maximum number of Kelvin-hours, with the bounds adapting online to meet it~\cite{shiUserspecifiedKelvinhourBudgets2026}, or as a bound on the average violation of a data-driven predictive controller~\cite{shiDisturbanceAdaptiveDataDrivenPredictive2026}.
Within RL a review of 79 studies found 68 reward structures with weighting coefficients chosen by trial~\cite{togashiRewardFunctionDesign2025}.
The weight is usually reported as a single value.

Comparisons of RL and MPC in building control differ in their ranking of the two methods.
On the BOPTEST case used here, an MPC on a reduced-order thermal model outperforms DDPG, DDQN and SAC~\cite{wangComparisonReinforcementLearning2023}.
For an equivalent problem, MPC outperforms model-free RL~\cite{arroyoComparisonOptimalControl2022}.
In residential demand response, a DQN policy reaches on average 62\,\% of the cost saving of MPC~\cite{kurteComparativeAnalysisModelfree2021}.
In commercial buildings, both RL and MPC hold the comfort constraints, however, RL saves less energy~\cite{ramanReinforcementLearningControl2020}.
Over a simulated year, SAC saves 7.1\,\% of energy, and white-, grey- and black-box MPC save 7.2 to 8.4\,\%~\cite{stoffelEvaluationAdvancedControl2023}.
In a field deployment in an occupied house, MPC saves 18.1\,\% of energy and model-based RL 20.9\,\%, with overlapping confidence intervals~\cite{mulayimComparativeFieldDeployment2027}.
On multi-energy systems with realistic, imperfect predictions, RL outperforms a linear MPC~\cite{ceustersModelpredictiveControlReinforcement2021}.
The advantage of MPC therefore depends on the accuracy of its model and its predictions.
Reviews of data-driven MPC and RL~\cite{zhangBuildingEnergyManagement2022} demonstrations~\cite{khabbaziLessonsLearnedField2025} report energy savings for both methods.
Each of these comparisons uses one formulation of comfort per method, either a weighted penalty or a constraint.
A measured difference between the methods therefore combines the effect of the method with the effect of the formulation.

A comparison of controllers requires the same plant and the same disturbances for each of them.
BOPTEST provides high-fidelity emulators of buildings for this purpose and is demonstrated with an economic MPC with soft comfort bounds on the BOPTEST \casename{} here~\cite{blumBuildingOptimizationTesting2021}.
BOPTEST-Gym makes its cases available to RL algorithms~\cite{arroyoOpenAIGymEnvironment2021}.

\section{PROBLEM DEFINITION}

Consider a discrete-time system
\begin{equation}
  x_{k+1} = f(x_k, u_k, d_k), \qquad y_k = h(x_k, u_k, d_k),
  \label{eq:system}
\end{equation}
with state $x_k$, input $u_k \in \mathcal{U}$, exogenous disturbance $d_k$ and output $y_k$.
The stage cost $\ell(y_k, u_k) \ge 0$ is to be minimized.
The constraint cost $c(y_k, u_k) \ge 0$ measures the violation of a constraint on the output and the input, and it is zero exactly when the constraint is satisfied.
The hard-constrained formulation is
\begin{equation}
  \min_{u_{0:K-1}} \; \mathbb{E}\Bigl[\textstyle\sum_{k} \ell_k\Bigr]
  \quad \text{s.t.} \quad c_k = 0 \;\; \forall k ,
  \label{eq:hard}
\end{equation}
with the expectation over the disturbance and the constraint required almost surely.
The soft-constrained formulation replaces the constraint by a weighted penalty,
\begin{equation}
  \min_{u_{0:K-1}} \; \mathbb{E}\Bigl[\textstyle\sum_{k} \ell_k + w\, c_k\Bigr], \qquad w > 0 .
  \label{eq:soft}
\end{equation}
In this paper, $\ell_k$ is the electrical energy of the heat pump in step $k$, defined in~\eqref{eq:stage}, and $c_k$ is the violation of the comfort bounds on the zone temperature, defined in~\eqref{eq:excess}.
The goal is to compare the energy use and the constraint violation that result from the two formulations, in MPC and in RL.

\section{METHODS}

\subsection{Reinforcement learning}
With the observed part of the disturbance added to the state, \eqref{eq:system} is a Markov decision process with transition kernel induced by $f$ and the disturbance~\cite{suttonReinforcementLearningIntroduction2018}.
The soft formulation~\eqref{eq:soft} becomes the reward $r_k = -\ell_k - w\, c_k$, and a policy $\pi$ maximizes $J_r(\pi) = \mathbb{E}_\pi[\sum_k \gamma^k r_k]$.
We use two off-policy actor--critic methods for continuous actions.
DDPG~\cite{lillicrapContinuousControlDeep2019} fits a critic $Q_\theta$ to the Bellman target and trains deterministic actor $\mu_\phi$ to maximize $Q_\theta(s, \mu_\phi(s))$, exploring by additive noise.
SAC~\cite{haarnojaSoftActorcriticOffpolicy2018} uses a stochastic actor, adds the policy entropy with a coefficient to the objective, and takes the minimum of two critics in the target~\cite{fujimotoAddressingFunctionApproximation2018}.

\subsection{Lagrangian reinforcement learning}
The hard formulation~\eqref{eq:hard} becomes a CMDP~\cite{altmanConstrainedMarkovDecision1999} with reward $r_k = -\ell_k$, cost $c_k$ and bound $d$,
\begin{equation}
  \max_\pi \; J_r(\pi)
  \quad \text{s.t.} \quad
  J_c(\pi) = \mathbb{E}_\pi\Bigl[\textstyle\sum_k \gamma_c^k c_k\Bigr] \le d .
  \label{eq:cmdp}
\end{equation}
It differs from~\eqref{eq:hard} in that it constrains an expected discounted sum instead of every step of every trajectory. 
Lagrangian methods solve~\eqref{eq:cmdp} through $\mathcal{L}(\pi, \lambda) = J_r(\pi) - \lambda\,(J_c(\pi) - d)$, $\lambda \ge 0$.
The policy is updated to maximize $\mathcal{L}$ at the current $\lambda$.
A second critic predicts $\hat{J}_c$, an estimate of $J_c$.
Then $\mathcal L$ is minimized with respect to the multiplier $\lambda$ using projected gradient descent
\begin{equation}
  \lambda \leftarrow \mathrm{clip}\bigl(\lambda + \eta\,(\hat{J}_c - d),\; 0,\; \lambda_{\max}\bigr) .
  \label{eq:dual}
\end{equation}
The multiplier increases while the estimated violation $\hat{J}_c - d$ is positive and decreases while it is negative.
The upper bound $\lambda_{\max}$ limits the weight that the multiplier can reach.
The Lagrange multiplier in the hard formulation~\eqref{eq:hard} has a similar effect to the weight in the soft formulation~\eqref{eq:soft}, with the difference that the multiplier is adjusted algorithmically by the update \eqref{eq:dual}. 
SAC-Lagrangian and primal-dual DDPG~\cite{liangAcceleratedPrimalDualPolicy2018} adapt the SAC and DDPG algorithm respectively by adding the cost critic and the update~\eqref{eq:dual}.

\subsection{Model predictive control}
At every step, MPC solves a finite-horizon problem using a model $\hat f, \hat h$ of~\eqref{eq:system}.
The problem starts from the state estimate $\hat x_k$ and uses a forecast $\hat d$ of the disturbance.
The controller applies the first input of the solution and repeats this at the next step~\cite{drgonaAllYouNeed2020}.
Both formulations use the problem
\begin{equation}
\begin{aligned}
  \min_{u} \quad & \textstyle\sum_{j=0}^{N-1} \ell(\hat y_{j+1}, u_j) + \rho\, c(\hat y_{j+1}, u_j) \\
  \text{s.t.} \quad & \hat x_{j+1} = \hat f(\hat x_j, u_j, \hat d_j), \quad \hat x_0 = \hat x_k, \\
  & \hat y_j = \hat h(\hat x_j, u_j, \hat d_j), \quad u_j \in \mathcal{U} .
\end{aligned}
\label{eq:mpc}
\end{equation}
For the soft formulation~\eqref{eq:soft}, the weight is $\rho = w$.
For the hard formulation~\eqref{eq:hard}, $\rho$ is set above the exact-penalty threshold~\cite{kerriganSoftConstraintsExact2000}.
Since the constraint cost $c$ increases linearly with the violation, the solution of~\eqref{eq:mpc} then equals the solution of the problem with the constraint $c = 0$ whenever that problem is feasible.
When it is infeasible, the solution minimizes the stage cost plus $\rho$ times the violation, and it approaches the solution with the least violation as $\rho$ grows~\cite{kerriganSoftConstraintsExact2000}.

\section{HVAC SYSTEM CONTROL}
\subsection{Plant}
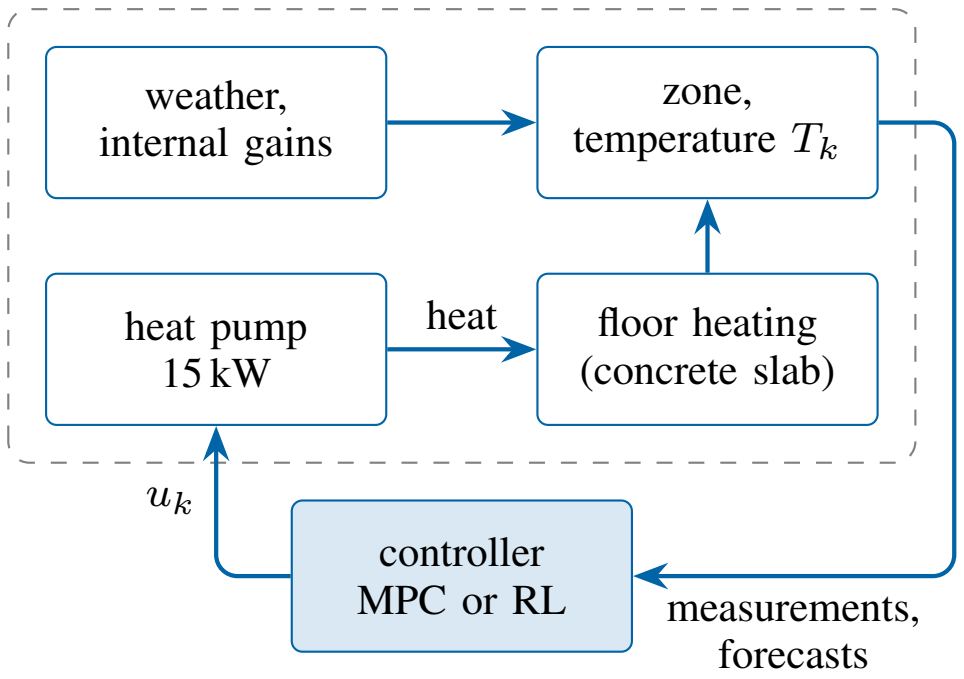
\begin{figure}[t]
  \centering
  \begin{tikzpicture}[font=\footnotesize,
      block/.style={draw=diagramblue, rounded corners=2pt, minimum width=2.25cm,
        minimum height=1cm, align=center, fill=white},
      arrow/.style={-{Stealth}, draw=diagramblue, thick}]
    \draw[gray, dashed, rounded corners] (-3,-1) rectangle (3,2.0);
    \node[block] (weather) at (-1.5-0.25/2, 1.25) {weather,\\internal gains};
    \node[block] (zone) at (1.5+0.25/2, 1.25) {zone,\\temperature $T_k$};
    \node[block] (pump) at (-1.5-0.25/2,-0.25) {heat pump\\15\,kW};
    \node[block] (floor) at (1.5+0.25/2,-0.25) {floor heating\\(concrete slab)};
    \node[block, fill=diagramblue!15] (controller) at (0,-1.75) {controller\\MPC or RL};
    \draw[arrow] (weather) -- (zone);
    \draw[arrow] (pump) -- node[above, black] {heat} (floor);
    \draw[arrow] (floor) -- (zone);
    \draw[arrow, rounded corners] (controller.west) -| node[pos=0.75, left, black] {$u_k$} (pump.south);
    \draw[arrow, rounded corners] (zone.east) -- ++(0.5,0) |- node[pos=0.75, below, black, align=center]
      {measurements,\\forecasts} (controller.east);
  \end{tikzpicture}
  \caption{The BOPTEST case \casename{} and the control loop. Every $\Delta t = 900$\,s the controller sets the compressor modulation $u_k$ from the measurements and the forecasts of the disturbance and of the comfort bounds.}
  \label{fig:plant}
\end{figure}

We use the BOPTEST case \casename{}~\cite{blumBuildingOptimizationTesting2021}.
It is a  single-zone residential building in Brussels with 192\,m$^2$ of floor area, heated by a 15\,kW modulating air-to-water heat pump through floor heating, without cooling.
The input is the compressor modulation $u_k \in [0, 1]$, held over a control step of $\Delta t = 900$\,s.
The disturbance $d_k$ is the weather and the internal gains of a family of five that occupies the building before 07:00 and from 20:00 on weekdays and throughout the weekend.
The BOPTEST case comes with one year of disturbance data.
The output is $y_k = (T_k, P^{\mathrm{hp}}_k, P^{\mathrm{fan}}_k, P^{\mathrm{pump}}_k)$, the zone operative temperature and the electrical power of the compressor, the evaporator fan and the emission pump.
The fan and the pump run when the compressor runs.
The energy of a step and of an episode, per floor area $A_{\mathrm f}$, are
\begin{equation}
  e_k = \tfrac{\Delta t}{A_{\mathrm f}}\bigl(P^{\mathrm{hp}}_k + P^{\mathrm{fan}}_k + P^{\mathrm{pump}}_k\bigr),
  \quad E = \textstyle\sum_k e_k\;[\mathrm{kWh/m^2}].
  \label{eq:energy}
\end{equation}

The interaction of the controller with the plant and its various inputs is visualized in Figure~\ref{fig:plant}.

\subsection{Comfort bounds and constraint cost}

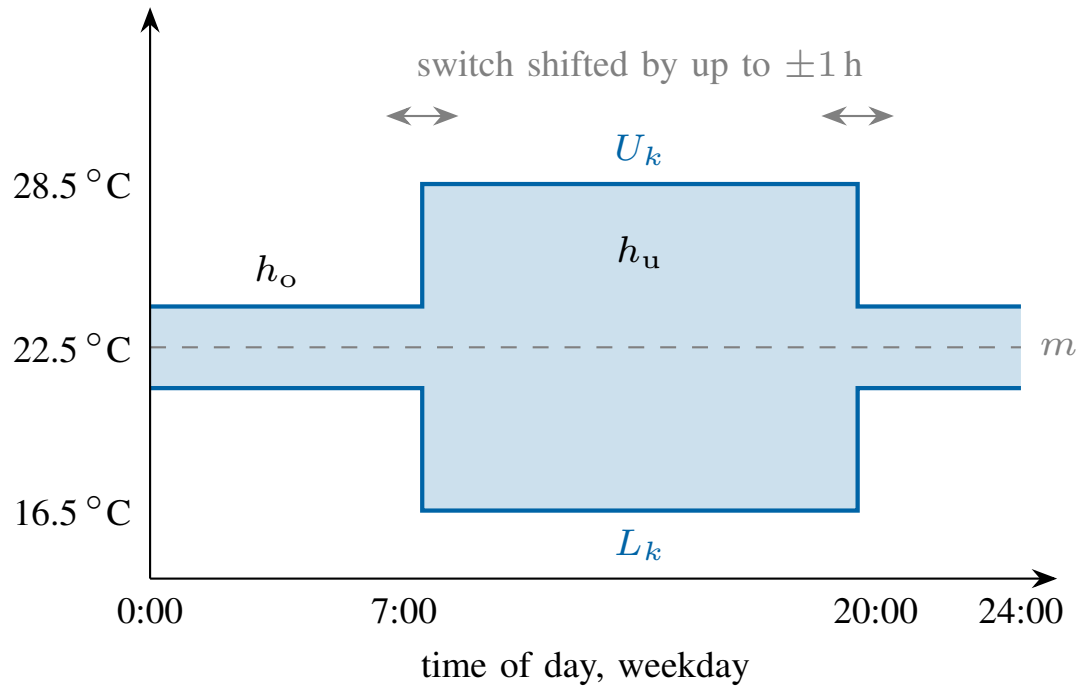
\begin{figure}[t]
  \centering
  \begin{tikzpicture}[x=0.24cm, y=0.18cm, font=\scriptsize]
    \fill[diagramblue!20] (0,7) -- (7.5,7) -- (7.5,2.5) -- (19.5,2.5)
      -- (19.5,7) -- (24,7) -- (24,10) -- (19.5,10)
      -- (19.5,14.5) -- (7.5,14.5) -- (7.5,10) -- (0,10) -- cycle;
    \draw[diagramblue, thick] (0,7) -- (7.5,7) -- (7.5,2.5) -- (19.5,2.5)
      -- (19.5,7) -- (24,7);
    \draw[diagramblue, thick] (0,10) -- (7.5,10) -- (7.5,14.5) -- (19.5,14.5)
      -- (19.5,10) -- (24,10);
    \draw[gray, dashed] (0,8.5) -- (24,8.5);
    \foreach \s in {7.5,19.5} \draw[gray, {Stealth}-{Stealth}] (\s-1,17) -- (\s+1,17);
    \node[gray, above] at (13.5,17.5) {switch shifted by up to $\pm 1$\,h};
    \draw[-{Stealth}] (0,0) -- (25,0);
    \draw[-{Stealth}] (0,0) -- (0,21);
    \foreach \h in {0,7,20,24} \node[below] at (\h,0) {\h:00};
    \foreach \c/\y in {16.5/2.5, 22.5/8.5, 28.5/14.5} \node[left] at (0,\y) {\c\,$^\circ$C};
    \node[diagramblue, above] at (13.5,14.5) {$U_k$};
    \node[diagramblue, below] at (13.5,2.5) {$L_k$};
    \node[gray, right] at (24,8.5) {$m$};
    \node at (3.5,11.3) {$h_{\mathrm o}$};
    \node at (13.5,12) {$h_{\mathrm u}$};
    \node[below] at (12,-2) {time of day, weekday};
  \end{tikzpicture}
  \caption{
    Visualization of comfort bounds for one week day.
    The comfort bounds for a weekday are centered around $22.5\,^\circ$C and defined by occupied and unoccupied half-widths $h_{\mathrm o} \in [0.5, 3]$K and $h_{\mathrm u}\in [3, 10]$K, resulting in lower and upper bounds $L_k$ and $U_k$.
    The switching times vary by plus or minus one hour from the default switching times at 07:00 and 20:00.}
  \label{fig:bounds}
\end{figure}

The bounds are centered on $m = 22.5\,^\circ$C, with a half-width $h_{\mathrm o}$ while the building is occupied and $h_{\mathrm u}$ otherwise,
\begin{equation}
    \begin{split}
    L_k &= m - h_k\\
    U_k &= m + h_k
    \end{split}
    \qquad
    h_k = \begin{cases} h_{\mathrm o} & \text{occupied} \\ h_{\mathrm u} & \text{otherwise} \end{cases}
\label{eq:bounds}
\end{equation}
In every episode, $h_{\mathrm o}$ and $h_{\mathrm u}$ are drawn log-uniformly from $[0.5, 3]$\,K and $[3, 10]$\,K respectively, and the switch times at 07:00 and 20:00 of each day are shifted by up to $\pm 1$\,h in steps of $\Delta t$, as shown in figure~\ref{fig:bounds}.
The switching times and bounds are sampled once per evaluation and test episode and are fixed for all controllers.

The constraint cost of a step and of an episode are
\begin{equation}
\begin{aligned}
  c_k &= \max(0, L_k - T_k) + \max(0, T_k - U_k) \;\; [\mathrm K], \\
  C &= \Delta t \textstyle\sum_k c_k \;\; [\mathrm{K\,h}] .
\end{aligned}
\label{eq:excess}
\end{equation}
The lower and the upper bound are combined into one constraint with one constraint cost.
Because $L_k < U_k$, at most one of the two terms of $c_k$ is positive at any step.
An episode can start outside the bounds, and the plant has no cooling.
Therefore $c_k=0$ does not hold at every step and~\eqref{eq:hard} is infeasible.
As a result, both methods must implement the hard formulation by a weight on the constraint cost that is large compared with the weights of the soft formulation.

\subsection{Formulations}
The stage cost is the energy of the step as a daily rate over $e_{\mathrm{ref}} = 0.5$\,kWh/m$^2$ per day,
\begin{equation}
  \ell_k = \frac{e_k}{e_{\mathrm{ref}}} \cdot \frac{24\,\mathrm h}{\Delta t},
  \label{eq:stage}
\end{equation}
so that $\ell_k$ and $c_k$ are of order one at a typical operating point.
The weighted learners receive the reward $r_k = -\ell_k - w\,c_k$, with $w$ on a grid of twelve values spaced logarithmically from 0.02 to 2.
The Lagrangian learners solve~\eqref{eq:cmdp} with $r_k = -\ell_k$, $\gamma_c = 0.99$, $d = 0$ and $\lambda_{\max} = 100$, so the comfort bounds are required at every step.
To compare the Lagrangian learners with the soft formulation at the weight their multiplier reaches, weighted SAC and DDPG are also trained at $w = \lambda_{\max} = 100$, with hyperparameters searched at this weight.
The MPC solves~\eqref{eq:mpc} with the same $\ell$ and $c$, and the weighted MPC uses $\rho = w$ on the same grid.
The constrained MPC uses $\rho = 1000$, which lies above the exact-penalty threshold (Section~\ref{sec:results_hard_soft}).

\subsection{Controllers}
\paragraph*{Learners}
A learner observes the vector
\begin{equation}
\begin{aligned}
  o_k = \bigl(&T_k - L_k,\; U_k - T_k,\; \tau_k,\; u_{k-1}, \\
  &T^{\mathrm{amb}}_{k+i},\; I^{\mathrm{gh}}_{k+i},\; L_{k+i},\; U_{k+i}\bigr)
\end{aligned}
\label{eq:observation}
\end{equation}
with 27 elements: the margins to both bounds, sine and cosine of the time of day and of the week $\tau_k$, the previous input, and the forecasts of the ambient temperature, the global horizontal irradiation and both bounds at $i \in \{0, 1, 3, 6, 12\}$\,h ahead.
Actor and critics have two hidden layers of 256 ReLU units.
The actor outputs $a_k \in \mathbb R$, and the input is $u_k = (\operatorname{clip}(a_k, -1, 1) + 1)/2$.
The critics are evaluated at the clipped output.
Because they then give no gradient for $|a_k| > 1$, the actor's loss contains the term $\kappa \max(0, |a_k| - 1 - \delta)^2$, with $\kappa$ and $\delta$ searched hyperparameters, which keeps the output near the range.
All learners are trained on episodes of 7 days.

\paragraph*{MPC}
The MPC is that of~\cite{blumBuildingOptimizationTesting2021}, with energy in place of energy cost.
Its model has the state $\hat x = (T, T^{\mathrm{mass}}, T^{\mathrm{wall}}, T^{\mathrm{floor}}, T^{\mathrm{cond}})$, temperatures connected by six resistances, and the disturbance $\hat d = (T^{\mathrm{amb}}, I^{\mathrm{dn}}, Q^{\mathrm{int}})$, the ambient temperature, the direct normal irradiation through a solar aperture and the internal gains.
The condenser heat and the compressor power are $u_j$ times factors affine in $T^{\mathrm{cond}}$ and $T^{\mathrm{amb}}$. 
The fan and pump draw $P^{\mathrm{fp}} = 531$\,W while a binary $b_j$ is one.
The specific optimization problem from~\eqref{eq:mpc} becomes
\begin{equation}
\begin{aligned}
  \min_{u, b, s} \quad & \textstyle\sum_{j=0}^{N-1} \bigl(p^{\mathrm{hp}}_j u_j + P^{\mathrm{fp}} b_j\bigr)\frac{24\,\mathrm h}{A_{\mathrm f}\, e_{\mathrm{ref}}} + \rho\, s_j \\
  \text{s.t.} \quad & \hat x_{j+1} = A_d \hat x_j + B_d\, q_j u_j + E_d \hat d_j, \quad \hat x_0 = \hat x_k, \\
  & L_{k+j+1} - s_j \le \hat T_{j+1} \le U_{k+j+1} + s_j, \quad s_j \ge 0, \\
  & 0.02\, b_j \le u_j \le b_j, \quad b_j \in \{0, 1\},
\end{aligned}
\label{eq:mpcfull}
\end{equation}

with $q_j$ and $p^{\mathrm{hp}}_j$ the condenser heat and compressor power per unit modulation, evaluated at the condenser temperature of the previous plan.
The slack $s_j$ makes the constraint cost~\eqref{eq:excess} linear: at the optimum it equals $c$ of the predicted zone temperature.
The problem is solved again when this changes the planned heat by more than 1\,\%.
Each problem is a mixed-integer linear program, solved by HiGHS to a gap of 1\,\% or a wall-clock limit of 2\,s.
An unscented Kalman filter~\cite{wanUnscentedKalmanFilter2000} estimates $\hat x_k$ from the measured zone temperature and, while the heat pump runs, the measured condenser heat.
We choose a fixed horizon of 24\, h, equivalent to $N=96$ steps at every weight.
The forecasts are exact.
The parameters are identified as in~\cite{blumBuildingOptimizationTesting2021} on two weeks of operation of the plant under the case's baseline PI controller with the case's own setpoints, and validated on the following two weeks.
Using the recorded modulation and disturbances, without state correction, the model reproduces the zone temperature with RMS errors of 0.14\,K on the fitting data and 0.25\,K on the validation data.

\paragraph*{Baseline PI controller}
The BOPTEST case includes a proportional--integral controller that sets the modulation to track a setpoint of the zone operative temperature~\cite{blumBuildingOptimizationTesting2021}.
We give it the setpoint $L_k + 0.2$\,K, the margin the case uses while occupied.
This is analogous to the baseline PI controllers' setpoint, which would otherwise be given by the case's internal static comfort bands.

\subsection{Training, Evaluation and Test Episodes}
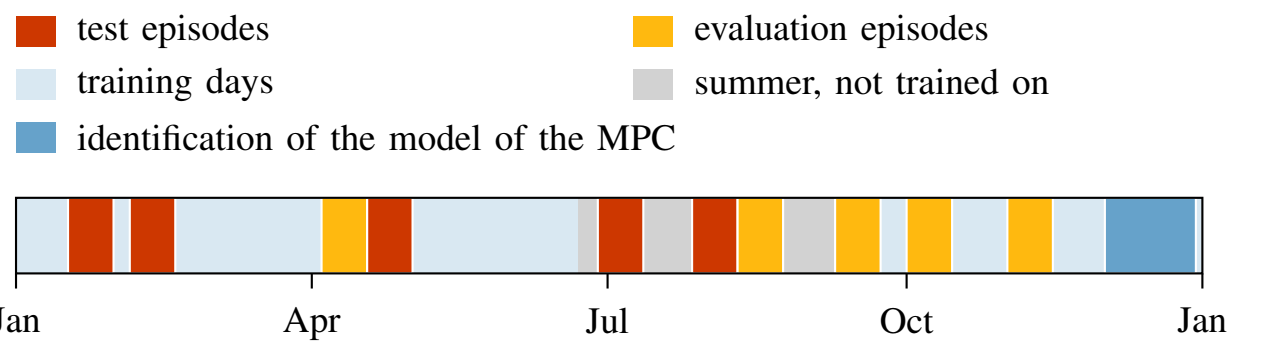
\begin{figure}[t]
  \centering
  \begin{tikzpicture}[x=0.0215cm, y=1cm, font=\scriptsize]
    \fill[diagramblue!15] (0,0) rectangle (365,0.5);
    \fill[gray!35] (173,0) rectangle (266,0.5);
    \foreach \a/\b in {16/30, 35/49, 108/122, 179/193, 208/222}
      \fill[testperiod, draw=white] (\a,0) rectangle (\b,0.5);
    \foreach \a/\b in {94/108, 222/236, 252/266, 274/288, 305/319}
      \fill[evaluationperiod, draw=white] (\a,0) rectangle (\b,0.5);
    \fill[diagramblue!60, draw=white] (335,0) rectangle (363,0.5);
    \draw (0,0) rectangle (365,0.5);
    \foreach \d/\m in {0/Jan, 91/Apr, 182/Jul, 274/Oct, 365/Jan} {
      \draw (\d,0) -- (\d,-0.1); \node[below] at (\d,-0.1) {\m};
    }
    \foreach \x/\y/\c/\l in {0/1.6/testperiod/test episodes,
        190/1.6/evaluationperiod/evaluation episodes,
        0/1.25/diagramblue!15/training days,
        190/1.25/gray!35/{summer, not trained on},
        0/0.9/diagramblue!60/identification of the model of the MPC} {
      \fill[\c] (\x,\y-0.1) rectangle (\x+12,\y+0.1); \node[right] at (\x+13,\y) {\l};
    }
  \end{tikzpicture}
  \caption{The days of the year: the ten episodes of 14 days, each after a warm-up of 7 days, the summer, and the four weeks on which the model of the MPC is identified and validated. Training episodes start on days outside the ten episodes and the summer.}
  \label{fig:periods}
\end{figure}

Separate evaluation and test episodes keep the hyperparameter search apart from the reported results. Ten episodes of 14 days, each preceded by a warm-up of 7 days, are fixed in advance: the two periods defined by the BOPTEST case and eight selected from weather statistics.
Five are evaluation episodes, used for every hyperparameter decision, and five are test episodes, used only for the reported results.
Training episodes start on random days outside all ten episodes and outside the summer, days 173 to 266, in which the heating-only plant needs almost no heating, as in~\cite{wangComparisonReinforcementLearning2023}. Because we are using a simulation where only heating is possible, we exclude the summer days where no heating is required across almost all days, which is consistent with prior work~\cite{wangComparisonReinforcementLearning2023}.
Results are reported on the three test episodes outside the summer.
Figure~\ref{fig:periods} shows how the year is split into training, test and evaluation episodes, where only periods outside of the summer season are used.

\subsection{Training}
The learners are trained on 128 environments in parallel on a JAX reimplementation of the case, both in the search and for the reported runs.
The reimplementation reproduces the episode sampling of the BOPTEST FMU and agrees with it to within 0.054\,K in zone temperature and 0.13\,\% in episode energy.
Each configuration is trained with five seeds for $1.2 \times 10^6$ transitions, and every evaluation, in the search and for the results, runs on the FMU.

\subsection{Hyperparameter search}
For each algorithm, and each weight of the weighted algorithms, we run a separate study of 128 trials to identify the best hyperparameters.
We use Optuna~\cite{akibaOptunaNextgenerationHyperparameter2019} with tree-structured Parzen Estimator (TPE) sampler, and asynchronous successive halving\cite{liSystemMassivelyParallel2020,eimerHyperparametersReinforcementLearning2023} pruner.
For the first 20 trials, the hyperparameters are chosen randomly.
Each trial trains three learners with different seeds, and is evaluated on the evaluation episodes after 1, 2 and $4 \times 10^5$ transitions, where it continues only if it is in the better half of the trials evaluated there.
Table~~\ref{tab:hyperparameters} lists the searched hyperparameters, their ranges and the chosen values.
\begin{table*}[t]
  \caption{Hyperparameters of the learners: the searched range and the chosen value. For the weighted learners, the chosen values are given as the range over the twelve weights of the grid and $w = 100$.}
  \label{tab:hyperparameters}
  \centering
  \setlength{\tabcolsep}{4pt}
  \begin{tabular}{llllll}
    \toprule
    & searched & SAC, weighted & DDPG, weighted & SAC-Lagrangian & primal-dual DDPG \\
    \midrule
    learning rate [$10^{-3}$] & 0.1--3, log & 0.18--2.8 & 0.5--3 & 2.9 & 3.0 \\
    discount factor & 0.97, 0.99, 0.995, 0.999 & 0.97--0.995 & 0.97--0.999 & 0.999 & 0.99 \\
    target update rate [$10^{-3}$] & 1--20, log & 1.4--19 & 1.2--20 & 1.1 & 2.8 \\
    batch size & 128, 256, 512 & 128--512 & 128--512 & 256 & 128 \\
    updates per transition & 1/8, 1/4, 1/2, 1 & 1/8--1 & 1/8--1 & 1 & 1 \\
    action-bound penalty $\kappa$ & 0.1--10, log & 0.1--6.8 & 0.1--9.5 & 0.11 & 0.14 \\
    action-bound margin $\delta$ & 0.05, 0.1, 0.2 & 0.05--0.2 & 0.05--0.2 & 0.05 & 0.2 \\
    initial entropy coefficient & 0.01--1, log & 0.012--0.87 & -- & 0.99 & -- \\
    target entropy & $-4$--0 & $-3.96$\,--\,$-0.0612$ & -- & $-0.174$ & -- \\
    exploration noise & 0.01--0.4, log & -- & 0.013--0.27 & -- & 0.021 \\
    multiplier learning rate [$10^{-3}$] & 0.1--100, log & -- & -- & 0.80 & 0.14 \\
    \bottomrule
  \end{tabular}
\end{table*}

A weighted algorithm is scored on $E \cdot 24\,\mathrm h / e_{\mathrm{ref}} + w\, C$, the episode sum of $(\ell_k + w\, c_k)\,\Delta t$, which is the quantity defined by its weight. A constrained algorithm is scored on $E$ subject to $C$ at most that of the baseline controller, with trials that exceed it ranked last by the amount they exceed it.
\subsection{Comparison of frontiers}
The frontier of controller B in an episode is the set of its non-dominated $(C, E)$ points and the straight lines between adjacent ones.
We count the points of controller A, over the weights and episodes, that this frontier dominates, that is, that some point of it has no higher $C$ and no higher $E$. A point on the frontier counts as dominated.
The count is over the weights from $0.107$ on, at which all weighted controllers heat.

\section{RESULTS AND DISCUSSION} 

\begin{table}[t]
  \caption{Energy $E$ and constraint cost $C$, as means over the three test episodes outside the summer.
  For the learners, mean and standard deviation over five seeds.
  $E$ in kWh/m$^2$, $C$ in K\,h.}
  \label{tab:main}
  \centering
  \setlength{\tabcolsep}{3pt}
  \begin{tabular}{llrr}
    \toprule
    & controller & $E$ & $C$ \\
    \midrule
    & baseline PI controller & 2.50 & 59.06 \\
    \midrule
    \multirow{5}{*}{\rotatebox{90}{soft}}
    & MPC, $w = 2$ & 2.16 & 6.47 \\
    & SAC, $w = 2$ & 2.11\,$\pm$\,0.03 & 13.06\,$\pm$\,5.86 \\
    & DDPG, $w = 2$ & 2.13\,$\pm$\,0.03 & 14.29\,$\pm$\,6.55 \\
    & SAC, $w = 100$ & 3.10\,$\pm$\,0.06 & 3.80\,$\pm$\,0.29 \\
    & DDPG, $w = 100$ & 2.83\,$\pm$\,0.10 & 5.17\,$\pm$\,0.39 \\
    \midrule
    \multirow{3}{*}{\rotatebox{90}{hard}}
    & MPC & 2.30 & 4.34 \\
    & SAC-Lagrangian & 2.81\,$\pm$\,0.03 & 7.49\,$\pm$\,3.31 \\
    & primal-dual DDPG & 2.76\,$\pm$\,0.10 & 12.65\,$\pm$\,6.02 \\
    \bottomrule
  \end{tabular}
\end{table}

\begin{figure}[t]
  \centering
  \includegraphics[width=\columnwidth]{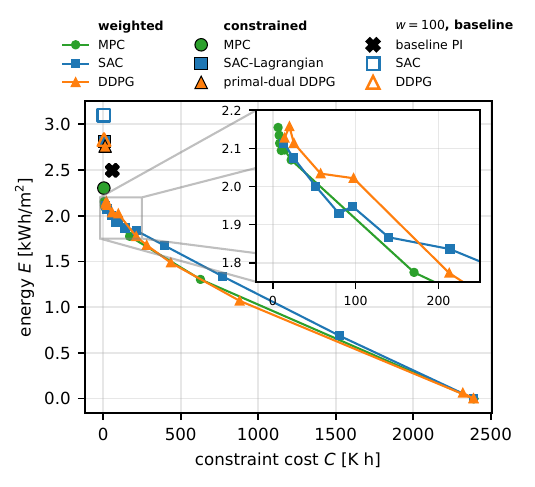}
  \caption{Energy $E$ against constraint cost $C$, test episodes outside the summer.
  Each weighted controller is a curve over its twelve weights.
  Each constrained controller is a filled point, and each weighted learner trained at $w = 100$ an open point.
  The inset enlarges the region $C \le 250$\,K\,h, $1.75 \le E \le 2.2$\,kWh/m$^2$.}
  \label{fig:frontier}
\end{figure}

\begin{figure}[t]
  \centering
  \includegraphics[width=\columnwidth]{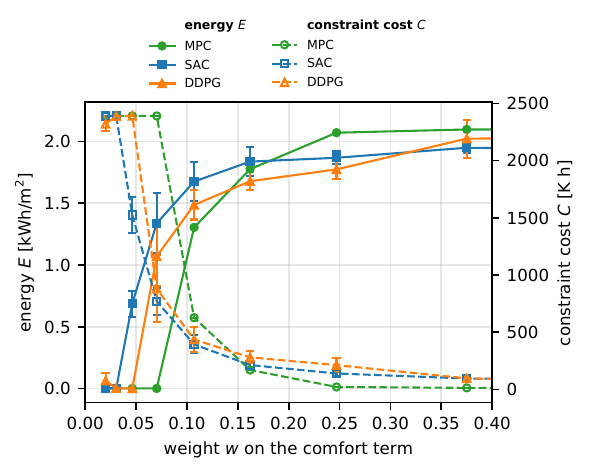}
  \caption{Energy $E$ (solid, linear left axis) and constraint cost $C$ (dashed, right axis) of the weighted controllers against the weight $w$, as means over the three test episodes outside the summer.
  The weights are spaced logarithmically, and the axis shows those up to 0.4.
  Error bars are the standard deviation over five seeds.}
  \label{fig:weight}
\end{figure}

\subsection{Hard against soft formulation} \label{sec:results_hard_soft}
In MPC the hard formulation uses 6.8\,\% more energy than the soft formulation at the largest weight and has a 33\,\% lower constraint cost (Table~\ref{tab:main}).
The constrained MPC reaches a lower constraint cost than the weighted MPC at every weight of the grid, and by the exact-penalty result~\cite{kerriganSoftConstraintsExact2000}, a larger weight brings the weighted MPC toward its operating point.

SAC-Lagrangian and primal-dual DDPG use 33\,\% and 30\,\% more energy than the weighted learner with the same base algorithm at $w = 2$, and have a 43\,\% and 11\,\% lower constraint cost.
With $d = 0$ and $c_k \ge 0$, the update~\eqref{eq:dual} only increases the multiplier, and the multiplier reaches $\lambda_{\max}$ in every run.
The difference in RL therefore depends on the choice $\lambda_{\max} = 100$. In MPC, $\rho = 1000$ is above the exact-penalty threshold, and from $\rho = 100$ to $1000$ the energy changes by 0.7\,\%. The two Lagrangian learners differ from each other by 1.7\,\% in energy, so the weight separates the learners more than the base algorithm does. Weighted DDPG trained at $w = 100$ uses 2.5\,\% more energy than primal-dual DDPG, which is within one standard deviation over seeds, and has a 7.48\,K\,h lower constraint cost (Table~\ref{tab:main}, Fig.~\ref{fig:frontier}).
SAC-Lagrangian uses 9.3\,\% less energy than weighted SAC at $w = 100$, at a 3.70\,K\,h higher constraint cost.

The hard controllers, and the weighted controllers from $w = 0.866$ on, hold the comfort bounds better than the baseline controller, since they receive forecasts of the comfort bounds and the baseline PI controller uses none. The baseline therefore does not heat in advance of the occupied bounds, and 29.3 of its 59.1 K h fall in the 2 h after a switch. The constrained MPC uses 8\,\% less energy than the baseline, and the Lagrangian learners 11 to 13\,\% more.

\subsection{Range of effective weights}
At low weights, not heating minimizes the weighted cost, and the constraint cost is 2390\,K\,h (Fig.~\ref{fig:weight}).
The weighted MPC starts to heat between $w = 0.0702$ and $0.107$, the learners at lower weights.
The MPC has no terminal cost, so its horizon of 24\,h limits how far ahead it trades energy against constraint cost.
Both energy and constraint cost vary most between $w \approx 0.05$ and $0.4$ for the weighted algorithms.
Fig.~\ref{fig:weight} shows that, for all algorithms, the energy rises and the constraint cost falls with increasing weight, and both change monotonically up to the run-to-run variation. 
At a fixed weight, the differences in $E$ and $C$ between the algorithms can be large, especially between $w = 0.02$ and $0.16$, since a given weight does not affect all algorithms equally. For each algorithm, however, some weight yields nearly the same $E$ and $C$, so their frontiers in Fig.~\ref{fig:frontier} lie close to each other. 
From $w = 0.375$ to 2, the energy of the weighted learners rises by 5 and 9\,\% while their constraint cost falls by a factor of about seven, and the energy of the weighted MPC rises by 3\,\% while its constraint cost falls by a factor of 2.2.
The slope of the frontier, the energy per Kelvin-hour of constraint cost avoided, therefore rises along it.
Between adjacent weights, it is 0.0004 to 0.003\,kWh/m$^2$ per K\,h for $w$ between 0.05 and 0.4, up to 0.034 between $w = 0.375$ and 2, and from $w = 2$ to the hard formulation 0.069 for MPC, 0.13 for SAC and 0.39 for DDPG.

\subsection{Choice of formulation}
The hard formulation uses 7\,\% more energy in MPC and about 30\,\% more in RL than the soft formulation at the largest weight.
The bounds cannot be held at every step.
Both methods therefore implement the hard formulation by a weight on the constraint cost that is large compared with the weights of the grid, $\rho = 1000$ in MPC and $\lambda_{\max} = 100$ in RL.
In MPC, the exact-penalty result makes this weight equivalent to the constraint.
In RL, the energy of the weighted learners at $w = 100$ lies closer to that of the Lagrangian learners than their energy at $w = 2$ (Table~\ref{tab:main}).
For a large $w$ and a large $\lambda_{\max}$, the operating points of the two formulations are similar.
The hard formulation is specified by the bounds, and the soft formulation additionally requires a value of the weight.
The energy use increases most with the weight between $w \approx 0.05$ and $0.4$.
Above $w \approx 0.4$, it continues to increase with the weight, at a much lower rate.
A single reported weight is informative only together with its position relative to this range.

\subsection{Constraint satisfaction}
All hard-constrained controllers violate the comfort bounds.
The constrained MPC leaves 4.34\,K\,h, and the Lagrangian learners leave 7.49 and 12.65\,K\,h, with a standard deviation over seeds of about half the mean.
The Lagrangian learners constrain only the expected discounted violation over the training distribution, and even this expectation remains positive: in all ten runs, the multiplier ends at its bound $\lambda_{\max} = 100$, with an estimated violation $\hat{J}_c - d$ between 1.8 and 5.3 for primal-dual DDPG and between 15.5 and 20.1 for SAC-Lagrangian.
\subsection{The Lagrangian learners}
Since an episode can start outside the comfort bounds, every policy has a positive expected violation, and the update~\eqref{eq:dual} raises the multiplier to its upper limit $\lambda_{\max} = 100$ in every run.
The Lagrangian learners thus end with a weight fifty times the largest weight of the grid, which explains their lower constraint cost and larger energy use compared with the weighted learners at $w = 2$. A positive bound $d$ would let the multiplier settle at a finite value, but it replaces the requirement at every step by a bound on the total violation, such as the Kelvin-hour limit of~\cite{shiUserspecifiedKelvinhourBudgets2026}, which is a different specification.
\subsection{Learners versus MPC}
At $w=2$, the weighted learners use about as much energy as the weighted MPC, 1 to 2 \% less, at twice its constraint cost.
From $w = 0.107$ on, the frontier of the weighted MPC dominates 12 of the 24 points of the weighted SAC and 17 of the 24 points of the weighted DDPG.
The frontiers of the learners dominate 7 and 3 of the 24 points of the MPC.
In the two cold test episodes, the frontier of the MPC dominates 9 and 13 of the 16 points of each learner, and the frontiers of the learners dominate 1 and 0 of its 16 points.
In the mildest test episode, the frontier of the weighted SAC dominates 6 of the 8 points of the MPC.
In the cold episodes, most operating points of a learner thus lie on or above the frontier of the weighted MPC. 
The MPC has more information than the learners, since a model-based controller needs an estimate of its state and forecasts of its disturbances.
Its filter estimates the hidden temperatures, and it receives forecasts of the internal gains over its whole horizon, while the learners have no past observations and receive forecasts at five lead times up to 12 h, without the internal gains.
Overall, the difference between the algorithms is small compared to the effect of the weight.

\subsection{Computational Cost}
\begin{table}[t]
  \caption{Computational cost of the controllers.}
  \label{tab:cost}
  \centering
  \setlength{\tabcolsep}{3pt}
  \begin{tabular}{lrrr}
    \toprule
    & Training & \multicolumn{2}{c}{Inference, one step} \\
    \cmidrule(lr){2-2} \cmidrule(lr){3-4}
    controller & [min] & mean [ms] & maximum [ms] \\
    \midrule
    MPC, $w = 2$ & 0.26 & 4035 & 6186 \\
    MPC, constrained & 0.26 & 4299 & 6283 \\
    SAC, $w = 2$ & 10.8 & 0.022 & 0.041 \\
    DDPG, $w = 2$ & 9.9 & 0.022 & 0.126 \\
    SAC-Lagrangian & 13.4 & 0.022 & 0.041 \\
    primal-dual DDPG & 10.7 & 0.022 & 0.126 \\
    \bottomrule
  \end{tabular}
\end{table}

Fitting the model of the MPC and training a learner are done once, and both can run on a computer other than the one in the building.
The controller in the building computes one control step at a time and stores its model or network (Table~\ref{tab:cost}).
Training is timed for one controller with its searched configuration, $1.2 \times 10^6$ transitions on one NVIDIA H100 NVL, and for the MPC as the fit of the model both MPCs share, on one core of an AMD EPYC 9454.
A control step is timed on one core of the same processor over the 1344 steps of one test episode, without the simulation.
Only the controller's computation is timed.
For a learner the normalization of its inputs, the network and the conversion of its output into a modulation, for the MPC the filter update, the assembly of the program and the runs of the solver.

A learner's step is the same at every step, a network of 72\,704 multiply--accumulates, and takes 22\,\textmu s.
The Lagrangian learners evaluate the same actor network as the weighted learner of their base algorithm.
The MPC solves a mixed-integer program with one binary per step of its horizon, whose worst-case cost grows exponentially with the horizon, up to three times per step.
Its mean step takes 4.0\,s weighted and 4.3\,s constrained.
The solver's limit of 2\,s per solve bounds the slowest step at 6.3\,s, at which the solver returns the best plan it has found.
Both fit the control step of 15\,min, but the learner's step is five orders of magnitude shorter and needs no solver.

\subsection{Limitations}
The study uses one case, one climate and one plant without cooling.
The forecasts are exact, and forecast error may affect the formulations differently.
The model of the MPC is identified on the simulation it controls.
Compared to the MPC, the learners don't get the internal gains or temperatures of the building mass and the condenser.
While the learners could learn to infer the internal gains, without past observations, the learners problem is partially observed.
The MPC estimates the temperatures with its filter.

The results of the MPC vary slightly between runs, because the solver stops at a wall-clock limit.
The learners are trained on a re-implementation of the BOPTEST test case, but every reported number and every search score comes from the BOPTEST emulator.

\section{CONCLUSION}
This paper isolates how the formulation of the comfort bounds and its weight determine the energy use of MPC and RL controllers in a high-fidelity building emulator.
At the largest weight of the grid, the hard formulation uses 7\,\% more energy than the soft formulation in MPC and about 30\,\% more in RL, at an 11 to 43\,\% lower constraint cost.

In both methods, the hard formulation reduces to a large weight on the constraint cost, the exact penalty in MPC and the upper limit of the Lagrange multiplier in RL.
In RL, the energy difference therefore depends on the upper limit chosen for the multiplier.
The energy use of the soft formulation depends on the weight mainly between $w \approx 0.05$ and $0.4$.
Above this range, the energy per avoided Kelvin-hour rises by about two orders of magnitude toward the hard formulation.
At the largest weight, the three algorithms differ by at most 2\,\% in energy, so the formulation and the weight affect the energy use more than the choice of algorithm.
Which algorithm reaches the better trade-off depends on the episode: the weighted MPC, which by design has more information, in the two cold test episodes, and the weighted SAC in the mildest one.

\section{ACKNOWLEDGEMENT}
The authors used Claude Opus 5.5 (Anthropic) to assist in writing text throughout all sections of this paper.
All AI-generated content was extensively reviewed, edited, and verified by the authors to ensure that the content is accurate.

\bibliography{references}

\end{document}